\documentclass[acmsmall,screen,authorversion]{acmart}
\AtBeginDocument{%
  }

\setcopyright{acmlicensed}
\copyrightyear{2018}
\acmYear{2018}
\acmDOI{XXXXXXX.XXXXXXX}
\acmConference[Conference acronym 'XX]{Make sure to enter the correct
  conference title from your rights confirmation email}{June 03--05,
  2018}{Woodstock, NY}
\acmISBN{978-1-4503-XXXX-X/2018/06}

\usepackage{algorithm}
\usepackage{multirow}
\usepackage{graphicx}
\usepackage{booktabs}
\usepackage{tikz}
\usepackage{multirow}
\usepackage{tikz}
\usepackage{xcolor}
\usepackage{subcaption}

\usepackage{enumitem}
\usepackage{booktabs}
\usepackage{listings}
\usepackage{threeparttable}
\usepackage{colortbl}%
\usepackage{makecell}
\usepackage{pifont} %
\usepackage{mdframed}
\usepackage{mathpartir}
\usepackage{mathtools}

\usepackage{xspace}
\usepackage{enumitem}
\usepackage{cuted}
\usepackage{adjustbox}
\usepackage{listings}
\usepackage{hyperref}
\usepackage[most]{tcolorbox}
\tcbuselibrary{skins,breakable}
\usepackage{caption}
\DeclareCaptionType{listing}[Listing][List of Listings]
\begin{document}

\title{Identifying Adversary Tactics and Techniques in Malware Binaries with an LLM Agent}
\newcommand{\todoc}[2]{{\textcolor{#1}{\textbf{#2}}}}
\newcommand{\todored}[1]{{\todoc{red}{\textbf{[#1]}}}}
\newcommand{\todogreen}[1]{\todoc{green}{\textbf{[#1]}}}
\newcommand{\todoblue}[1]{\todoc{blue}{\textbf{[#1]}}}
\newcommand{\todoorange}[1]{\todoc{orange}{\textbf{[#1]}}}
\newcommand{\todobrown}[1]{\todoc{brown}{\textbf{[#1]}}}
\newcommand{\todogray}[1]{\todoc{gray}{\textbf{[#1]}}}
\newcommand{\todopink}[1]{\todoc{pink}{\textbf{[#1]}}}
\newcommand{\todopurple}[1]{\todoc{purple}{\textbf{[#1]}}}
\newcommand{\todo}[1]{\todored{TODO: #1}}

\newcommand{\name}{{\textsc TTPDetect}\xspace}
\newcommand{\renamer}{renaming agent }
\newcommand{\tnum}{203}
\newcommand{\subtnum}{453}
\newcommand{\zx}[1]{\todoorange{ZX: #1}}
\newcommand{\xz}[1]{\todored{XZ: #1}}
\newcommand{\xx}[1]{\todoblue{XX: #1}}
\newcommand{\ly}[1]{\todogreen{YL: #1}}
\newcommand{\mz}[1]{\todobrown{MZ: #1}}
\newcommand{\cp}[1]{\todopurple{CP: #1}}

\newtcolorbox{querybox}[1]{%
  enhanced,
  breakable,
  sharp corners,
  colback=gray!10,
  colframe=gray!60,
  colbacktitle=gray!30,
  coltitle=black,
  fonttitle=\scriptsize\bfseries,
  title={#1},
  boxrule=0.6pt,
  boxsep=0.5mm,
  left=1mm,right=1mm,top=1mm,bottom=1mm,
  before upper=\parindent0pt\scriptsize,
}

\author{Zhou Xuan}
\affiliation{%
  \institution{Purdue University}
  \city{West Lafayette}
  \country{USA}}

\author{Xiangzhe Xu}
\affiliation{%
  \institution{Purdue University}
  \city{West Lafayette}
  \country{USA}}

\author{Mingwei Zheng}
\affiliation{%
  \institution{Purdue University}
  \city{West Lafayette}
  \country{USA}}

\author{Louis Zheng-Hua Tan}
\affiliation{%
  \institution{Purdue University}
  \city{West Lafayette}
  \country{USA}}

\author{Jinyao Guo}
\affiliation{%
  \institution{Purdue University}
  \city{West Lafayette}
  \country{USA}}

\author{Tiantai Zhang}
\affiliation{%
  \institution{Purdue University}
  \city{West Lafayette}
  \country{USA}}

\author{Le Yu}
\affiliation{%
  \institution{Purdue University}
  \city{West Lafayette}
  \country{USA}}

\author{Chengpeng Wang}
\affiliation{%
  \institution{Purdue University}
  \city{West Lafayette}
  \country{USA}}

\author{Xiangyu Zhang}
\affiliation{%
  \institution{Purdue University}
  \city{West Lafayette}
  \country{USA}}

\renewcommand{\shortauthors}{Trovato et al.}

\begin{abstract}
Understanding TTPs (Tactics, Techniques, and Procedures) in malware binaries is essential for security analysis and threat intelligence, yet remains challenging in practice. Real-world malware binaries are typically stripped of symbols, contain large numbers of functions, and distribute malicious behavior across multiple code regions, making TTP attribution difficult. Recent large language models (LLMs) offer strong code understanding capabilities, but applying them directly to this task faces challenges in identifying analysis entry points, reasoning under partial observability, and misalignment with TTP-specific decision logic.

We present \name, the first LLM agent for recognizing TTPs in stripped malware binaries. \name combines dense retrieval with LLM-based neural retrieval to narrow the space of analysis entry points. \name further employs a function-level analyzing agent consisting of a Context Explorer that performs on-demand, incremental context retrieval and a TTP-Specific Reasoning Guideline that achieves inference-time alignment. We build a new dataset that labels decompiled functions with TTPs across diverse malware families and platforms. \name achieves 93.25\% precision and 93.81\% recall on function-level TTP recognition, outperforming baselines by 10.38\% and 18.78\%, respectively. When evaluated on real world malware samples, \name recognizes TTPs with a precision of 87.37\%. For malware with expert-written reports, \name recovers 85.7\% of the documented TTPs and further discovers, on average, 10.5 previously unreported TTPs per malware.

\end{abstract}



\received{20 February 2007}
\received[revised]{12 March 2009}
\received[accepted]{5 June 2009}

\authorsaddresses{}
\maketitle

\section{Introduction}
Malware has become one of the most pervasive and damaging cyber threats in recent years. Recent statistics show that aggregated reported direct losses from cyber incidents since 2020 have reached nearly \$28 billion, while total direct and indirect costs are estimated to account for 1\%–10\% of global GDP\cite{cyber_risk}. Malware based cyber-attacks remain the most prominent threat, with a broad reach and a significant financial impact\cite{EURSPOL}.

To defend against malware, the security community has developed a wide range of malware analysis techniques, including static analysis\cite{FlowDroid, Apposcopy, ASTROID}, dynamic analysis \cite{ndroid, betterpmp}, and learning-based methods \cite{fdroid, semanticmodeling, MalFocus, humanvsml, malstudy, API2Vec, guidedretrain}. LLM-based methods \cite{lamd, MalLoc} have emerged as a promising direction, leveraging LLM's rich external knowledge and strong reasoning capabilities. However, the majority of existing work focuses on identifying malware from benign software (i.e., identification) and classifying malware into specific families (i.e, attribution) \cite{Drebin, zhang_ccs14, TESSERACT, FeatureSmith}. 
Such analyses provide limited insight into which parts of the code implement malicious behavior or what concrete actions the malware performs. Beyond detection, effective defense against malware requires understanding how an attack operates, specifically what concrete adversarial behaviors are implemented and how they are composed within the malware.

The MITRE ATT\&CK framework \cite{mitre} provides a standardized, behavior-centric taxonomy of adversary tactics and techniques (TTPs), capturing the actions an adversary performs. Mapping malware behavior to TTPs is a critical task in cyber threat intelligence (CTI), enabling analysts to reason about adversarial intent, compare threats across campaigns, and prioritize mitigation strategies. For example, a stealer malware may perform Process Injection (T1055) to execute within a legitimate process, followed by Exfiltration Over C2 Channel (T1041) to transmit stolen data. Identifying such behavior sequences in terms of TTPs provides a higher-level, interpretable view of an attack, bridging low-level implementation details with actionable defensive insights.

Despite its importance, recognizing TTPs directly from malware binaries remains challenging. In real-world settings, malware binaries are typically stripped of symbols and debugging information, making original function and variable names unavailable and complicating analysis \cite{debin, DroidScope}. Existing TTP mapping approaches primarily operate on artifacts that provide richer semantic context, such as CTI reports \cite{report_ttp1}, script files \cite{LADE} or interpreted malware packages \cite{genttp}. Other work leverages malware execution traces \cite{tracettp1} or system-level logs\cite{trec}, but these dynamic approaches are heavyweight and costly to scale. These approaches are less applicable when only stripped binaries are available.

Large language models (LLMs) have recently demonstrated strong capabilities in code understanding and reasoning~\cite{DBLP:journals/corr/abs-2507-05269,rfcaudit,DBLP:conf/sp/ZhengXZ25,DBLP:conf/icml/Guo0XS025}, motivating their application to malware analysis. However, applying LLMs to TTP recognition in stripped binaries introduces several challenges. First, malware binaries often contain hundreds or thousands of functions, far exceeding the input length limitations of current LLMs \cite{lamd} and forcing analysis to operate at the function level. A single function, however, rarely provides sufficient evidence for accurate TTP attribution, leading to partial observability \cite{POMDP}. Second, LLMs are not inherently aligned with the structured, domain-specific reasoning required for TTP analysis and may overgeneralize based on superficial cues, resulting in hallucinations. Third, when extending function-level analysis to the binary level, it is difficult to identify analysis entry points, namely which functions in a large binary should be analyzed for which TTPs. The combinatorial space of possible candidate function–TTP pairs creates a severe scalability challenge, making naïve exhaustive analysis computationally infeasible.

In this paper, we propose \name, the first LLM agent for recognizing TTPs in stripped malware binaries. \name mirrors how human analysts perform TTP attribution, in which analysts first rapidly narrow down where to focus their analysis and then carefully examine contextual evidence, applying TTP-specific reasoning before making a final attribution. First, \name combines dense retrieval and an LLM-based neural retrieval to narrow the space of candidate function–TTP pairs, addressing the binary-level challenge of identifying analysis entry points. Second, for each candidate pair, \name applies a function-level TTP analyzing agent consisting of a Context Explorer for on-demand incremental context retrieval to mitigate partial observability and a TTP-Specific Reasoning Guideline to alleviate misalignment. 

We further introduce a new dataset that provides function-level TTP annotations for stripped binaries across diverse malware families and platforms. Using this dataset, we demonstrate that \name substantially outperforms baseline prompting approaches in both precision and recall. When evaluated on real world malware samples, \name not only recovers TTPs documented in expert-written reports but also discovers previously unreported TTPs that are validated as true positives through manual analysis.

In summary, this work makes the following contributions:

\begin{itemize}
  \item We propose \name, the first LLM agent that recognizes TTPs in stripped malware binaries at both the function and binary levels.
  \item We propose a hybrid retrieval strategy that combines dense retrieval with LLM-based neural retrieval to narrow the space of candidate function–TTP pairs, thereby identifying promising analysis entry points. We further design a function-level analyzing agent consisting of a Context Explorer that performs on-demand, incremental context retrieval, and a TTP-Specific Reasoning Guideline that achieves inference-time alignment.
  \item We build a new dataset that labels decompiled functions with TTPs across diverse malware families and platforms.
  \item \name achieves 93.25\% precision and 93.81\% recall on function-level TTP recognition, outperforming baselines by 10.38\% and 18.78\%, respectively. When evaluated on real world malware samples, \name recognizes TTPs with a precision of 87.37\%. For latest malware with expert-written reports, \name recovers 85.7\% of the documented TTPs and further discovers, on average, 10.5 previously unreported TTPs per malware.

\end{itemize}

\section{Motivation}
\label{sec:motivation}

\begin{figure*}[t]
    \centering
    \begin{subfigure}[t]{0.49\linewidth}
        \centering
        \includegraphics[width=\linewidth]{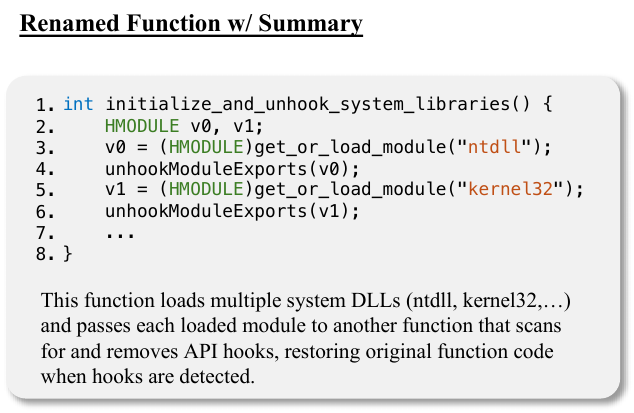}
        \caption{Example function and summary}
        \label{fig:gen_moti1}
    \end{subfigure}
    \hfill
    \begin{subfigure}[t]{0.49\linewidth}
        \centering
        \includegraphics[width=\linewidth]{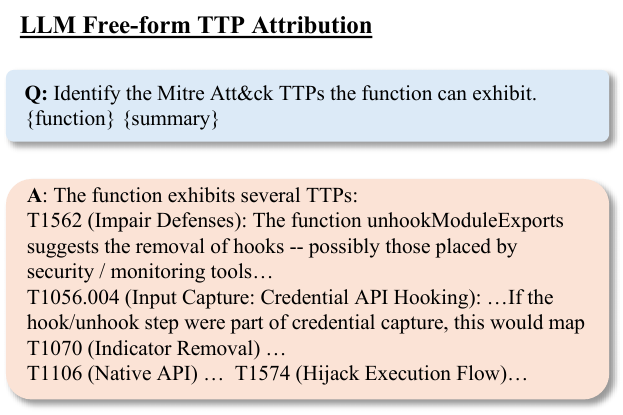}
        \caption{LLM free-form TTP attribution result}
        \label{fig:gen_moti2}
    \end{subfigure}

    \caption{An example of LLM Free-form TTP Attribution (Claude~3.7~Sonnet). The input is a renamed decompiled function with summary. Q denotes the input prompt and A denotes the LLM’s generated response. This function exhibits T1562 (Impair Defenses). The LLM correctly recognizes T1562, however it produces 4 false positives. }
    \label{fig:motivation}
\end{figure*}

\begin{figure*}
    \centering

    \begin{subfigure}{\linewidth}
        \centering
        \includegraphics[width=\linewidth]{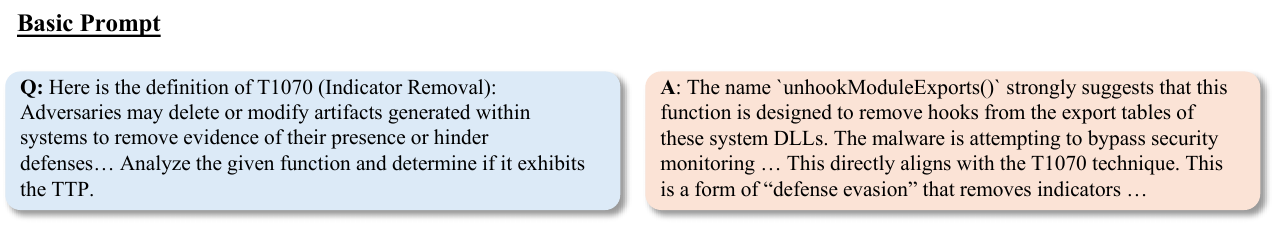}
        \caption{TTP presence prediction on whether the function in the motivation example exhibits T1070. The LLM still produces a false positive.}
        \label{fig:basic_prompt}
    \end{subfigure}

    \begin{subfigure}{\linewidth}
        \centering
        \includegraphics[width=\linewidth]{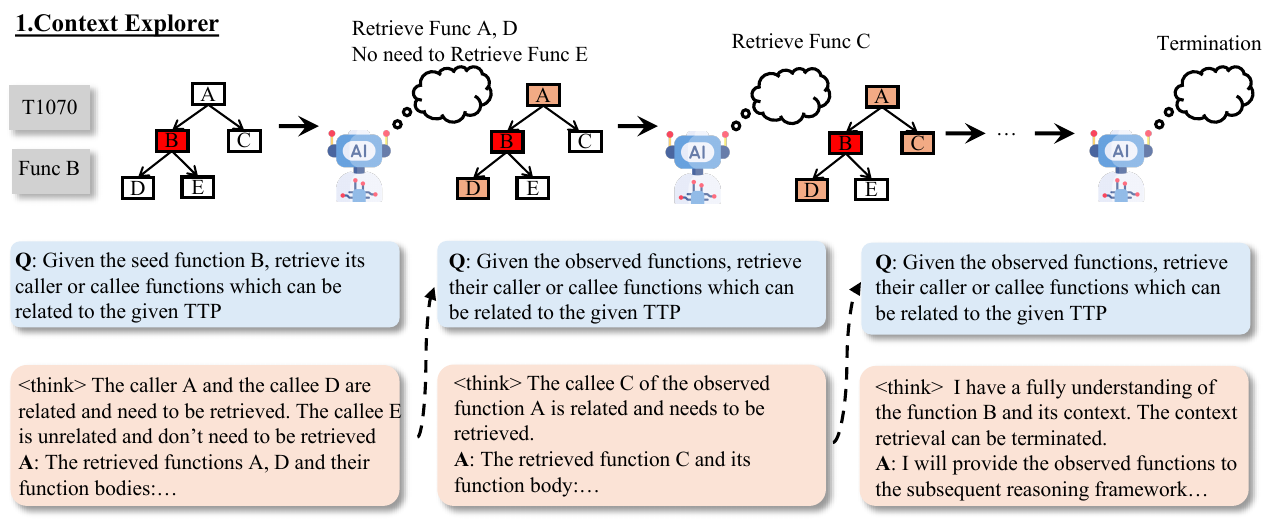}
        \caption{Context Explorer. The function names are simplified as A, B, C, D and E, which denote initialize\_api\_and\_resume\_execution, initialize\_and\_unhook\_system\_libraries, splice\_function\_code, unhookModuleExports, MapFileToReadOnlyMemory, separately. \textless think \textgreater denotes the internal thinking on which functions to retrieve. }
        \label{fig:moti_agentA}
    \end{subfigure}

    \begin{subfigure}{\linewidth}
        \centering
        \includegraphics[width=\linewidth]{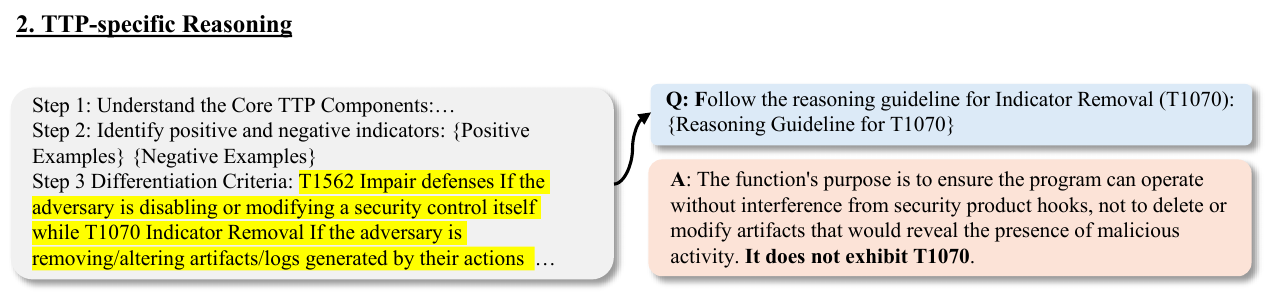}
        \caption{TTP-specific Reasoning. The left side is the pre-synthesized TTP-specific reasoning guideline.}
         \label{fig:moti_agentB}
    \end{subfigure}

    \caption{ Comparison of the Basic Prompt and the Analyzing Agent for TTP presence prediction}
    \label{fig:motivation_agent}
\end{figure*}

\subsection{Motivating Example}
\label{subsec:motivation_example}
Fig.~\ref{fig:motivation} presents a motivating example adapted from a Windows malware binary. 
Fig.~\ref{fig:gen_moti1} shows a decompiled function after function renaming, where address-based identifiers in the stripped binary are replaced with semantically meaningful function names to facilitate analysis.
For example, the original identifier \texttt{sub\_4016A0} is renamed to \texttt{unhookModuleExports} at line~4.
By removing API hooks installed by defensive software (e.g., Trusteer Rapport\cite{rapport}), the malware disables security monitoring and evades detection. This behavior corresponds to the TTP
T1562 (Impair Defenses), which involves disabling or modifying defensive mechanisms in a victim environment. Correctly identifying such TTPs is crucial for malware analysis and threat intelligence, as TTP-level attribution provides insight into adversary behavior and supports subsequent investigation and mitigation.

\subsection{Limitations of Existing Work}
Although TTP recognition is critical for malware analysis, existing work rarely addresses mapping stripped binaries to TTPs. MalEval~\cite{llm_malware_behavior} proposes an evaluation framework for auditing Android malware behaviors, but its labels are dataset-specific rather than aligned with standardized taxonomies such as ATT\&CK TTPs. It identifies pre-defined behavior including stealing, mining, but it could not attribute the malware to the 
comprehensive TTP set.
Prior research on TTP attribution primarily targets artifacts that provide richer semantic context, including cyber threat intelligence (CTI) reports~\cite{report_ttp1} and interpreted malware packages~\cite{genttp}. The text, source code or metadata are unavailable to facilitate the analysis of our motivation example. 
Other techniques rely on dynamic execution traces~\cite{tracettp1} or system logs~\cite{trec}. While effective in some settings, collecting such signals requires heavyweight execution and is prone to failure in practice, as malware can evade dynamic analysis using logic bombs or environment checks\cite{forceexe}. The motivating example in Fig.~\ref{fig:motivation} performs defense impairment only after specific runtime conditions are met (e.g., execution within a protected browser process). As a result, the relevant behavior may not be triggered during sandboxed execution, causing trace- or log-based approaches to miss the TTP entirely.
This gap motivates a static, context-aware approach that operates directly on stripped binaries and explicitly reasons about TTPs.

\subsection{Challenges of Using LLMs}
\label{subsec:challenges}
To address the limitations of existing work, we apply LLMs because they have stronger capabilities in general code understanding and reasoning. 
Providing the entire binary to an LLM is infeasible because real-world malware binaries often contain thousands of functions, far exceeding context window limits and degrading generation quality even for long-context models~\cite{lamd, long_context_code}. On the other hand, due to safety alignment constraints, LLMs may refuse to analyze malware when the complete malware code is provided as input\cite{openaisafety, claudesafety}.
Therefore, a feasible alternative is to perform function-level analysis, following prior work~\cite{deepreflect, FGGat, Maltracker} and standard human analyst practice~\cite{human_re}.

An intuitive method is to directly prompt an LLM with a renamed decompiled function and ask it to enumerate the TTPs that the function exhibits, as shown in Fig.~\ref{fig:gen_moti2}. This approach depends on the model to freely enumerate candidate TTPs and iteratively assess each candidate to determine the final TTP set. However, free-form generation is inherently unreliable, as both candidate selection and candidate verification can be misaligned with the decision criteria needed for precise TTP attribution.
In this example, the LLM correctly identifies T1562 (Impair Defenses) but also produces four false positives (T1056.004, T1070, T1106, and T1574).

Free-form generation alone is often insufficient for complex tasks that require multi-step reasoning, leading to overgeneralization and hallucination, whereas explicitly structuring the task can mitigate the weaknesses of free-form generation \cite{lost_in_middle, researchrubrics}. A straightforward idea is to decompose TTP attribution into a candidate suggestion stage that proposes plausible TTPs and a binary prediction stage that verifies each candidate. However, this decomposition alone does not fully address the limitations of free-form generation.
Fig.~\ref{fig:basic_prompt} illustrates the TTP-specific verification of T1070 for the motivating function~(Fig.~\ref{fig:gen_moti1}), where the model is provided with the TTP description to guide a binary decision on TTP presence, yet still produces a false positive. While the model may correctly recognize that the function removes hooks from system DLLs, it can overgeneralize from surface cues (e.g., “removal”) and map the behavior to a wrong TTP such as T1070 (Indicator Removal). 
Having examined function-level TTP attribution, we next consider how to recognize TTPs at the binary level. A natural extension is to apply this function-level procedure to every function in the binary and then aggregate the predicted TTPs across functions. In practice, however, real-world binaries contain hundreds or thousands of functions, and for each function, the analysis must first identify a set of candidate TTPs and then perform TTP presence prediction for each function–TTP pair. As a result, binary-level attribution requires evaluating a large number of function–TTP combinations, making naïve extensions computationally expensive and difficult to scale.

Recognizing TTPs from stripped malware binaries using LLMs poses challenges at both the binary and function levels. At the binary level, the large number of functions and the broad TTP space make it difficult to identify which function–TTP pairs are worth detailed analysis. At the function level, attribution is hindered by partial observability, since relevant behavior is frequently realized through interactions among multiple functions rather than a single isolated function, and by reasoning misalignment, where LLMs overgeneralize from surface cues and confuse semantically adjacent TTPs, as illustrated by the motivating example.

\noindent{\bf Challenge 1 (Binary-Level):  Difficulties in Identifying Analysis Entry Points.}
We define an analysis entry point as a candidate function–TTP pair that warrants detailed inspection. In practice, a malware binary may contain hundreds or thousands of functions, many of which are benign utilities, compiler-generated helpers, or library wrappers. In the motivating example, the number of functions increases by roughly 10 times from source code to the stripped binary. As a result, the few functions responsible for T1562 (defense impairment) are obscured among a large number of irrelevant functions. Meanwhile, the ATT\&CK framework defines over 200 TTPs. As a result, the vast majority of candidate function–TTP pairs are irrelevant to actual TTP implementation. Blindly evaluating all pairs with an LLM is therefore prohibitively expensive. An effective approach must instead provide a lightweight mechanism to identify a small set of promising candidate function–TTP pairs as analysis entry points before invoking any costly function-level analysis.

\noindent{\bf Challenge 2 (Function-Level): Partial Observability in Stripped Functions.} Even after selecting a candidate function–TTP pair as the analysis entry point, accurately determining whether the function exhibits the TTP is non-trivial when only the function body is provided in isolation. In malware, behavior is often distributed across multiple functions, and analysis starting from an isolated function induces a partially observable environment~\cite{POMDP}. Without additional context, the LLM cannot reliably determine how a function is used or what objective it ultimately serves.

Both caller and callee context can be essential. In our dataset, we observe a function that extracts browser cookies and stores the collected data in a file. When the LLM is asked whether this function exhibits T1074 (Data Staged), it answers negatively, reasoning that adversaries typically stage collected data prior to exfiltration and that, based on the function alone, there is no evidence that the stored data is later used for exfiltration. This example illustrates how partial observability at the function level can lead to false negatives when the broader execution context is missing.
Callee context is equally important. Note that although recovered callee names are visible in the function body, these identifiers, whether tool-generated, ML-predicted, or even developer-assigned, are often insufficiently precise\cite{Context2Name, symlm, varbert, gennm} and rarely capture the full semantic meaning. As a result, the model may over-interpret an imprecise name such as \texttt{unhookModuleExports} and speculate about behaviors that are not actually present (e.g., credential API hooking in T1056.004 as shown in Fig.~\ref{fig:gen_moti2}), especially when the relevant implementation details reside in deeper callees that are not shown.
Simply including all direct and transitive callees does not work in practice, as the call graph is often very large such that a caller may have numerous callees with many utility functions that do not provide useful hints.

\noindent{\bf Challenge 3 (Function-Level): Misalignment on TTP Reasoning.} 
Even with sufficient context, LLMs may still misattribute TTPs because their reasoning does not align with the domain-specific logic required for TTP analysis. In our example, the LLM correctly infers that the function removes API hooks, but incorrectly associates this behavior with T1070 (Indicator Removal). Because both T1070 involves some notion of “removal,” the model overgeneralizes and selects the wrong TTP. This error stems from misaligned reasoning traces. LLMs are not trained on TTP-specific decision processes and lack an internal structure that mirrors how security analysts reason about TTP boundaries, necessary conditions, and adversarial intent. As a result, generated rationales often rely on superficial cues, incomplete logical chains, or hallucinated intermediate steps, leading directly to incorrect final decisions. As the right part of Fig.~\ref{fig:basic_prompt} shows, the LLM’s final reasoning step makes an incorrect evidential link. It treats removing defensive hooks as evidence of indicator removal, even though T1070 requires actions that remove or manipulate forensic artifacts to hinder detection.

\begin{figure*}[h]
    \centering

    \includegraphics[width=0.95\linewidth]{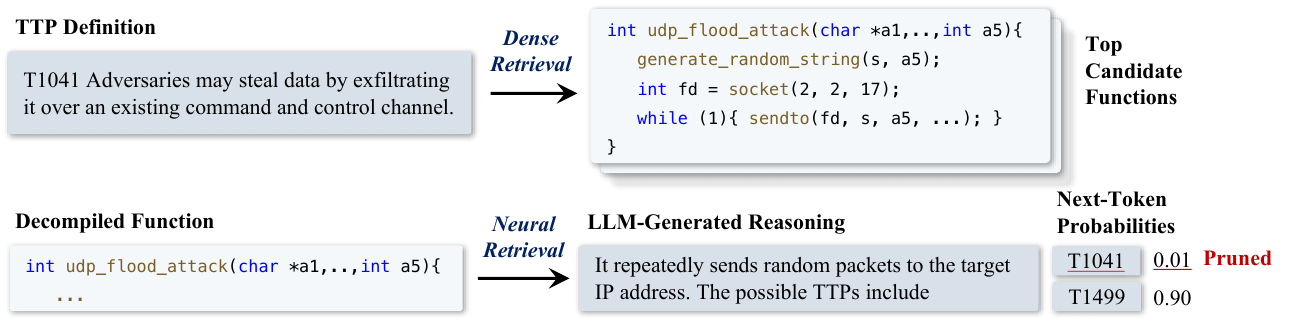}
    \caption{ Hybrid retrieval. Given the TTP definition of T1041, the dense retriever retrieves multiple candidate functions, ranking udp\_flood\_attack as a top candidate. In neural retrieval, the LLM is prompted with the decompiled function to generate an over-inclusive set of potentially relevant TTPs. T1041 is pruned.
    }
    \label{fig:moti_retrieval}
\end{figure*}

\subsection{Our Method}

We are inspired by how human analysts work when recognizing TTPs in binaries. They first rapidly narrow down where to focus their analysis, and then carefully examine contextual evidence, applying TTP-specific knowledge before making a final attribution. We design our system to mirror this process.
At a high level, our approach decomposes TTP recognition into two stages. First, we apply lightweight retrieval to quickly narrow the space of plausible function–TTP pairs, addressing the binary-level challenge of identifying analysis entry points. Second, for each candidate pair, we apply a function-level TTP analysis agent consisting of a Context Explorer for on-demand incremental context retrieval to mitigate partial observability and a TTP-Specific Reasoning Guideline to alleviate misalignment.

\noindent{\bf Solution for Challenge 1: Hybrid Retrieval of Candidate Function–TTP Pairs} 
To identify promising analysis entry points at the binary level, we adopt a hybrid retrieval strategy that combines dense retrieval and neural retrieval. Dense retrieval embeds TTP descriptions and function representations into a shared vector space and retrieves candidate functions based on cosine similarity. However, dense retrieval is not always reliable. As shown in Fig.~\ref{fig:moti_retrieval}, when using the description of T1041 (Exfiltration Over C2 Channel) as the query, the dense retriever returns multiple candidate functions, with \texttt{udp\_flood\_attack} ranked as a top candidate despite being incorrect. This behavior arises because the embedding model primarily attends to salient structural signals, such as networking-related operations, rather than TTP-specific semantics. In neural retrieval, the LLM is prompted with the decompiled function to generate an over-inclusive set of potentially relevant TTPs. This design prioritizes recall, aiming to reduce missed TTPs caused by insufficient context or reasoning misalignment. TTPs that remain irrelevant even under this permissive setting can then be pruned in later stages.

Next, we discuss how to address the function-level challenges. Fig.~\ref{fig:moti_agentA} and~\ref{fig:moti_agentB} illustrate how our solutions jointly correct the false positives produced by the basic prompt in Fig.~\ref{fig:basic_prompt}.

\noindent{\bf Solution for Challenge 2:  Context Explorer} To address partial observability, we introduce a Context Explorer that mirrors how a human analyst incrementally gathers evidence around a function of interest. Rather than analyzing a function in isolation, the explorer selectively expands along the call graph to retrieve only context that is relevant to understanding the function’s behavior.
As illustrated in Fig.~\ref{fig:moti_agentA}, the process begins from a seed function (Function~B). The agent iteratively retrieves caller or callee functions whose recovered identifiers suggest potentially meaningful behavior, while pruning those that appear to be generic utilities or unrelated implementation. This selective expansion avoids exhaustive traversal of the call graph and focuses the analysis on semantically relevant code.

In the motivating example, the agent first retrieves Functions~A and~D, while pruning Function~E, whose name suggests a generic memory-mapping utility. Inspection of Function~D confirms that it restores original API instructions, indicating defensive impairment. Further inspection of Function~A reveals a call to Function~C, which performs code injection, confirming malicious intent. The exploration terminates once the agent determines that the collected context is sufficient, and the retrieved functions are passed to the subsequent reasoning module.

\noindent{\bf Solution for Challenge 3: Inference-Time Alignment} 
To reduce reasoning misalignment, we introduce an inference-time alignment strategy that constrains the model to follow structured, TTP-specific reasoning. Training-time alignment is impractical due to the scarcity of high-quality TTP reasoning traces and the high cost of manual annotation. Therefore, we inject structured guidance only at inference time. For each TTP, we pre-synthesize a reasoning
guideline from the ATT\&CK knowledge base. The TTP-specific guideline consists of core
TTP components, positive and negative indicators, and differentiation criteria. As shown in Fig.~\ref{fig:moti_agentB}, the agent applies these guidelines to generate structured reasoning traces. In the example, By explicitly walking through the reasoning guideline for T1070, the LLM distinguishes indicator removal from defense impairment, as the highlighted part shows. Conditioned on this structure, the agent correctly rejects T1070 and avoids the false positive produced by the basic prompt.

\section{Problem Formulation}\label{sec:formulation}

We study the problem of recognizing TTPs from stripped malware binaries at the function level.
Let $F_{\text{bin}}=\{f_1,\ldots,f_n\}$ denote the set of functions recovered from a binary, and let
$T=\{t_1,\ldots,t_m\}$ denote the set of ATT\&CK TTPs under consideration.
Our goal is to determine, for each function–TTP pair $(f,t)$, whether function $f$ implements the TTP $t$.

We formulate the process of binary-level TTP attribution as a factorized policy:
\begin{equation}
\pi( \cdot \mid F_{\text{bin}}, T ),
\end{equation}
which models the overall probability of assigning TTPs to functions given the binary.

As shown in Eq.~\ref{eq:all}, the policy decomposes into three stages.
First, the retrieval stage prunes the space of possible function--TTP pairs.
For each TTP $t$, a dense retriever produces a candidate function set $\mathcal{R}_k(t) \subseteq F_{\text{bin}}$.
Here, $\mathcal{R}_k(t)$ denotes the top-$k$ functions retrieved for TTP $t$, where $k$ is a hyperparameter that limits the number of candidate functions considered per TTP.
In addition, we require the TTP likelihood predicted for a function to exceed a threshold, i.e., $P(t \mid f) > \tau$.
Only function--TTP pairs that satisfy both conditions are retained for subsequent exploration and reasoning.

Second, the exploration stage models the selection of contextual functions.
Given a candidate function $f$, the policy $\pi_{ctx}(F_{ctx}\mid f)$ retrieves a set of related functions $F_{ctx}$ (e.g., callers, callees, or semantically related functions) that may provide additional behavioral evidence.

Finally, the reasoning and decision stage determines whether a TTP applies to a function.
In this stage, we first generate a TTP-specific reasoning guideline $g$ according to $\pi_g(g \mid t)$, which encodes high-level behavioral criteria and domain knowledge associated with TTP $t$.
Conditioned on the explored context $F_{ctx}$, the TTP $t$, and the generated guideline $g$, the policy $\pi_{reason}(v \mid F_{ctx}, t, g)$ produces a binary decision $v \in \{\texttt{true}, \texttt{false}\}$.
Only pairs with $v=\texttt{true}$ are retained as positive TTP attributions.

This formulation explicitly captures the multi-stage nature of function-level TTP recognition and highlights how retrieval, context exploration, and reasoning jointly address the challenges of identifying analysis entry points, partial observability, and misalignment.

\begin{equation}
\label{eq:all}
\begin{aligned}
\pi( \cdot \mid F_{\text{bin}}, T ) &=
\prod_{\substack{f \in F_{\text{bin}} \\ t \in T}}
\underbrace{
\mathbb{I}\!\left(f \in \mathcal{R}_k(t)\right)\;
\mathbb{I}\!\left(P(t \mid f) > \tau\right)
}_{(1)\ \text{Retrieve}}
\\
&\quad
\underbrace{\pi_{ctx} ( F_{ctx} \mid f )}_{(2)\ \text{Explore}}\;
\underbrace{
\pi_g(g \mid t)\;
\pi_{reason}(v \mid F_{ctx}, t, g)\;
\mathbb{I}(v=\texttt{true})
}_{(3)\ \text{Reason \& Decision}}
\end{aligned}
\end{equation}

\section{Design}

\begin{figure*}[h]
\centering
\includegraphics[width=\linewidth]{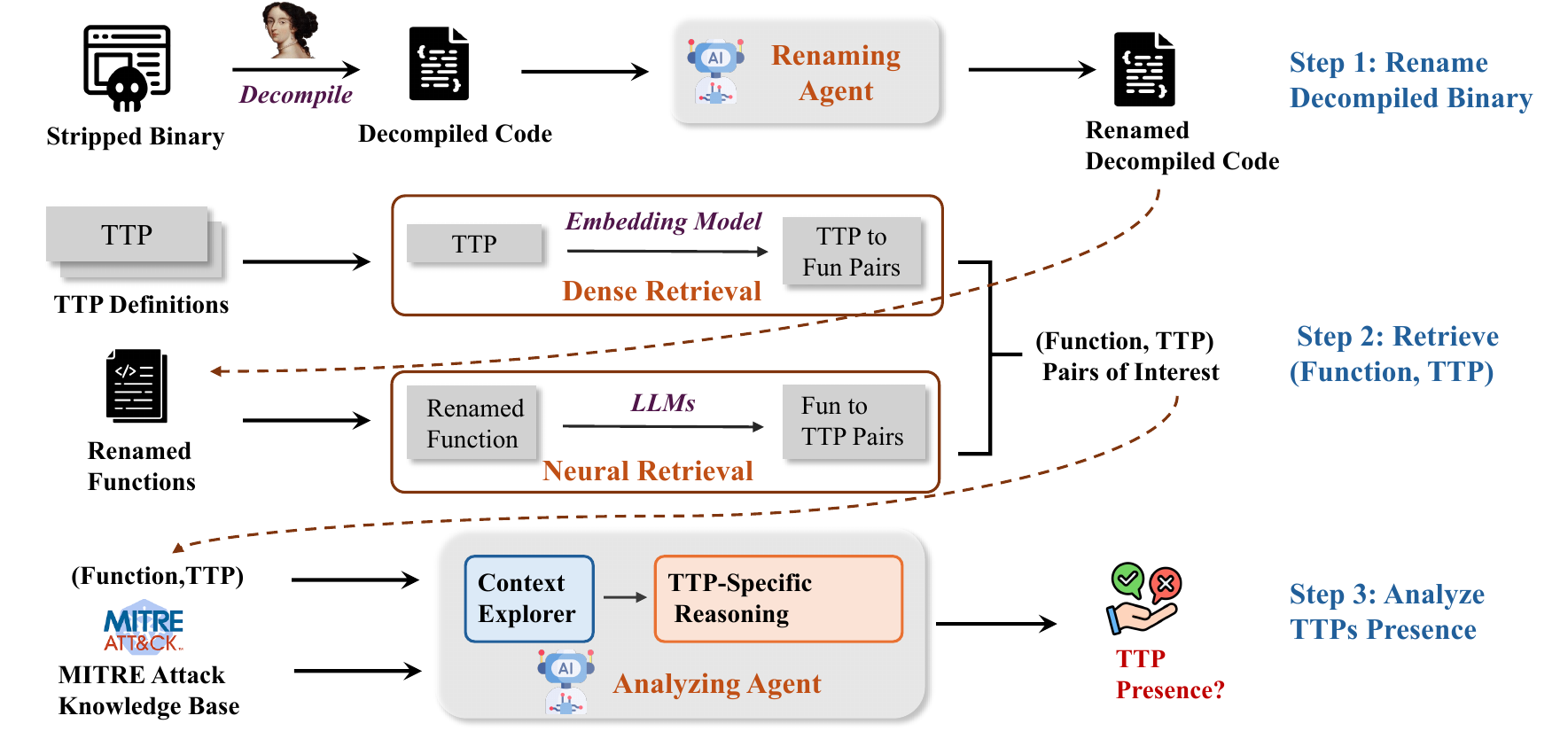}
\caption{\name Overview. 
}
\label{fig:overview}
\end{figure*}

As shown in Fig.~\ref{fig:overview}, \name follows a three-step pipeline. First, \name renames decompiled functions with semantically meaningful identifiers using an LLM-based agent. Second, it narrows the set of candidate (function, TTP) pairs through a combination of dense and neural retrieval. Finally, for each candidate pair, an analyzing agent selectively explores program context and applies TTP-specific reasoning to determine whether the TTP is present.

\subsection{Step 1: Decompiled Binary Renaming}
\label{sec:renamer}

As mentioned in Sec.~\ref{subsec:motivation_example}, stripped binaries use address-based identifiers (e.g., \texttt{sub\_401000}) for functions, which provide little information about their behavior and make analysis difficult. Recent work on stripped binaries has shown that LLMs can recover meaningful symbol names when given limited program context~\cite{gennm, ReSym}. We adopt a
similar idea by introducing a LLM-based renaming agent to assign descriptive, human-readable names to decompiled functions. For each target function, the agent analyzes the decompiled function body together with brief summaries of its direct callees, and generates a concise summary and a corresponding function name that reflects the function’s behavior. We restrict the context to direct callees, as incorporating deeper call-chain context would significantly increase the analysis scope and lead to context explosion.
The LLM prompt used for function renaming is shown in Listing~\ref{lst:rename-prompt}.

Functions are renamed in a bottom-up order over the call graph, ensuring the required callee summaries are available.
Functions that do not call other functions are processed first using only their own decompiled code. Functions that call others are processed afterward, once summaries for all called functions have been generated.
Cycles in the call graph are handled by temporarily assigning placeholder summaries and revisiting the affected functions after additional context has been accumulated. This process produces a renamed binary that supports more effective downstream analysis.

\begin{figure}[t]
\centering

\begin{minipage}[t]{0.39\linewidth}
\begin{querybox}{Query for Renaming}
\scriptsize
\noindent\textbf{Below is your code snippet.}\\[0.5ex]
\textit{$\langle$code $\rangle$}\\[0.8ex]

\noindent\textbf{Question:} You will be given the \textit{function body with callee names recovered} and
\textit{summaries of the direct callees}. Please analyze the function and provide:
\begin{enumerate}[leftmargin=*]
  \item Function summary (1--3 sentences).
  \item Recovered function name.
\end{enumerate}
\end{querybox}
\vspace{-3mm}
\captionsetup{type=listing}
\caption{LLM prompts used for renaming.}
\label{lst:rename-prompt}
\end{minipage}
\hfill
\begin{minipage}[t]{0.59\linewidth}
\begin{querybox}{Query for TTP Retrieval}
\scriptsize
\noindent\textbf{Below is your code snippet.}\\[0.5ex]
\textit{$\langle$code$\rangle$}\\[0.8ex]

\noindent\textbf{Question:} You will be given the \textit{renamed function body} and
\textit{its summary}. Please analyze the function and identify an over-inclusive set of potentially relevant ATT\&CK TTPs that this function exhibits. For each TTP, provide:
\begin{enumerate}[leftmargin=*]
  \item Brief reasoning (2--3 sentences).
  \item Confidence score (0.0--1.0).
\end{enumerate}
\end{querybox}
\vspace{-3mm}
\captionsetup{type=listing}
\caption{LLM prompts used for TTP retrieval.}
\label{lst:TTP-prompt}
\end{minipage}

\end{figure}
\vspace{-1mm}

\subsection{Step 2: Function-TTP Pair Retrieval}
\label{subsec:retrieval}
After obtaining a renamed binary, we identify candidate function-TTP pairs . To identify a small set of candidate pairs, we propose a two-stage retrieval pipeline that combines embedding-based dense retrieval with LLM-based neural retrieval.

\noindent{\bf Dense Retrieval} performs embedding-based retrieval to obtain an initial set of function--TTP candidate pairs. We encode each ATT\&CK TTP using its textual definition and each function using its renamed decompiled code. Both are embedded into a shared vector space using a code-capable embedding model (e.g., OpenAI text-embedding-3-large\cite{text-embedding-3-large}). For each TTP, we retrieve the top-\(k\) most similar functions based on cosine similarity, where \(k\) is a hyperparameter that limits the number of candidate functions considered per TTP. This stage is designed to maximize recall, producing a coarse candidate set that may still include many irrelevant pairs.

\noindent{\bf Neural Retrieval} applies an LLM to further narrow the candidate set.
As shown in Listing~\ref{lst:TTP-prompt}, for each function, the model is provided with its renamed decompiled code and brief summaries of its callee functions. Based on this information, the LLM generates an over-inclusive set of potentially relevant TTPs that the function may exhibit, together with brief reasoning and confidence scores. This permissive generation enables further pruning while preserving recall, as TTPs that are clearly incompatible with the function are discarded at this stage.
As illustrated in Fig.~\ref{fig:moti_retrieval}, the malware binary does not exhibit T1041 (Exfiltration Over C2 Channel) in any function. However, when iterating over all TTPs, dense retrieval may still retrieve a function implementing a DDoS attack for T1041 due to superficial similarity in network-related operations. In contrast, it is unlikely for the LLM to overgeneralize a DDoS function to exfiltration behavior. This filtering does not require deep contextual exploration or fine-grained reasoning, yet effectively removes clearly implausible function–TTP pairs. It additionally prunes candidates using confidence scores produced by the LLM, discarding pairs whose scores fall below a predefined threshold. This pruning step further reduces the number of candidates, thereby decreasing the need to invoke the more expensive function-level analyzing agent.

\subsection{Step 3: TTP Presence Analysis}

\begin{figure*}[t]
    \centering

    \begin{subfigure}{0.95\linewidth}
        \centering
        \includegraphics[width=\linewidth]{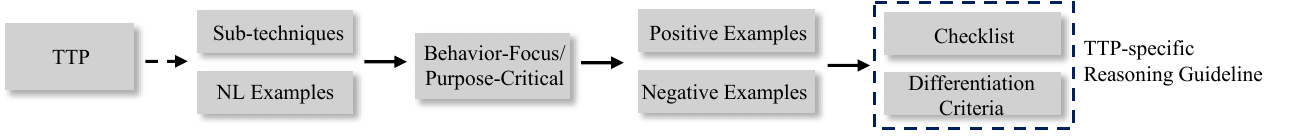}
        \caption{Workflow for reasoning guideline synthesis}
        \label{fig:reason}
    \end{subfigure}

    \begin{subfigure}{0.95\linewidth}
        \centering
        \includegraphics[width=\linewidth]{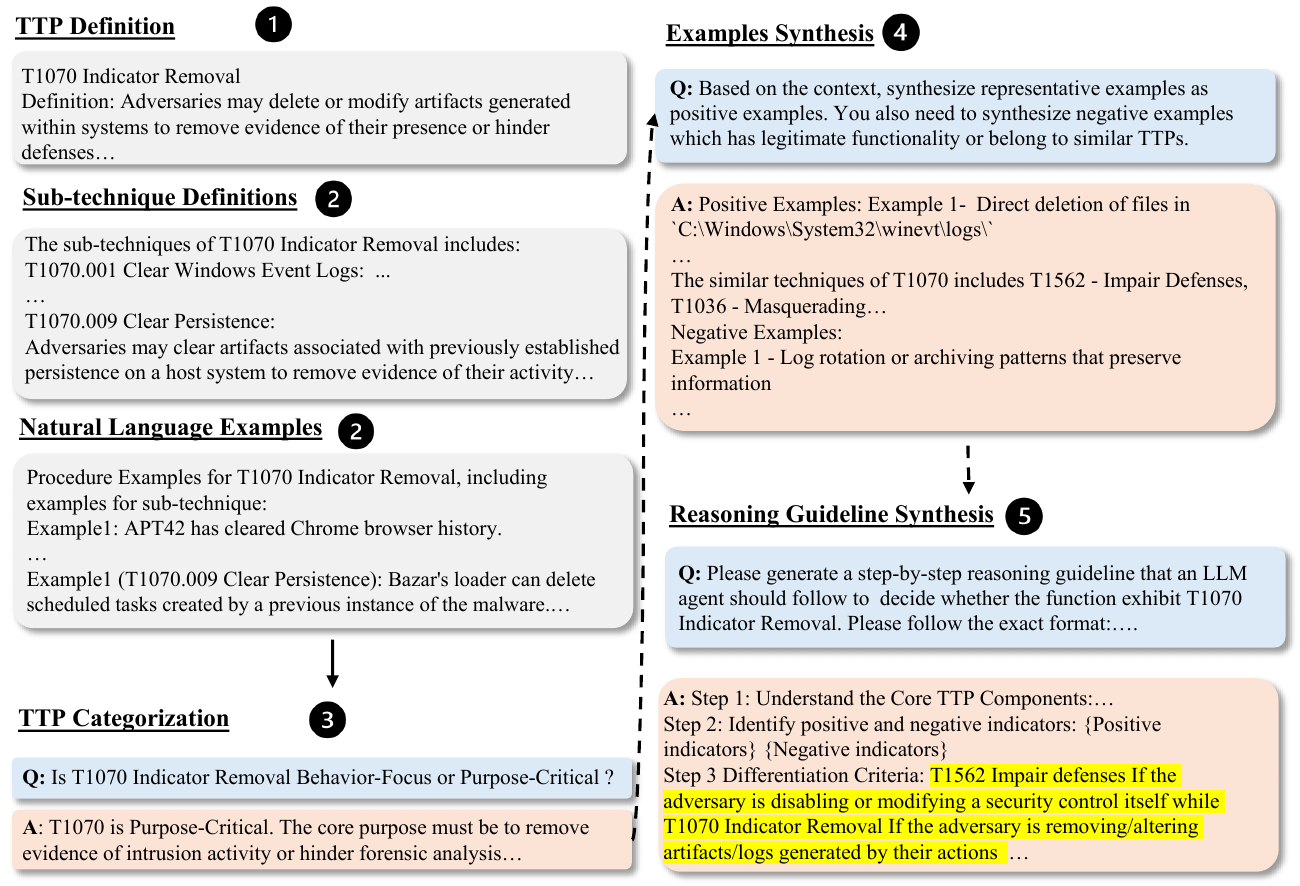}
        \caption{An example (T1070) for reasoning guideline synthesis. The number corresponds to different stages in (a). }
         \label{fig:reason_case}
    \end{subfigure}

    \caption{ Reasoning guideline synthesis}
    \label{fig:reason}
\end{figure*}

For each function–TTP pair retained after Step~2, \name applies the analyzing agent to determine whether the function implements the target TTP.
The agent first conducts demand-driven context exploration that selectively retrieves relevant program context, followed by TTP-specific reasoning guideline that guides the LLM to evaluate the collected context to make the final decision.

\subsubsection{Context Exploration} 

Given a target function, the Context Explorer incrementally retrieves only the program context needed to assess the TTP. Following the practice of \cite{DBLP:conf/icml/Guo0XS025, rfcaudit, multiagentfaultlocalization}, the Analyzing Agent is equipped with two primitive retrieval tools: (a) \texttt{retrieve\_function(func\_name)}, which returns the code of the specified function; and (b) \texttt{retrieve\_caller(func\_name)}, which returns the caller functions of the given function. Starting from the function itself, the agent selectively fetches callers and callees using the tools. At each step, it decides whether additional context is necessary based on the behavioral requirements of the target TTP. This targeted exploration avoids irrelevant utility code while preserving all information needed for accurate attribution.

\subsubsection{TTP-Specific Reasoning}
\label{subsubsec:design_alignment}
After collecting sufficient context, the agent determines whether the code meets the requirements of the target TTP. To mitigate misalignment, \name uses TTP-specific reasoning guidelines distilled offline from the MITRE ATT\&CK knowledge base and reused across all binaries at inference time.

Fig.~\ref{fig:reason} illustrates how \name constructs a structured reasoning guideline for each ATT\&CK TTP.
At step~1, we extract the formal definition of the TTP from the knowledge base. At step~2, we augment this definition with associated sub-technique and natural-language procedure examples. Sub-techniques broaden the space of possible implementations, while procedure examples describe how the TTP manifests in real attacks. Although these examples do not include source code, they provide detailed behavioral descriptions that modern LLMs can effectively relate to binary-level observations.
Notably, different TTPs require different attribution logic. Some are behavior-focused, where the presence of a characteristic action is sufficient regardless of intent. For example, Process Discovery(T1057) applies whenever code enumerates processes, whether through APIs such as \texttt{EnumProcesses}, scanning \texttt{/proc}, or querying task lists, independent of whether the intent is benign or malicious. Other TTPs are intent-critical, where similar low-level actions may be benign unless accompanied by evidence of adversarial purpose. For instance, Data Staged (T1074) involves aggregating data in preparation for exfiltration. Benign data aggregation alone is insufficient without indications of subsequent transmission to an external destination. At step~3, we prompt the LLM to classify the TTP using the accumulated context. The classification result is then reused as additional context in step~4, where the LLM synthesizes representative positive examples that capture common implementations of the TTP across platforms, APIs, and malware families, as well as complementary negative examples that reflect benign behaviors or closely related TTPs. By contrasting positive and negative cases, the synthesis clarifies which signals are essential and which are misleading.
Finally, at step 5, the LLM consolidates all information into a concise reasoning guideline that specifies the required components, indicative behaviors, and differentiation criteria. During inference, this guideline serves as a structured checklist that constrains the model’s reasoning, ensuring that each TTP is evaluated using consistent, TTP-specific criteria.

\section{Evaluation}

We implement \name atop AutoGen v0.6.1 and use Claude~3.7~Sonnet as the underlying LLM for all experiments. We use IDA Pro 7.6 to generate decompiled code from stripped binaries. For dense retrieval, we adopt the OpenAI \texttt{text-embedding-3-large} model to generate embeddings and use FAISS~v1.11.0 to build the retriever. We use MITRE ATT\&CK~v16.1 as the knowledge base. We evaluate the effectiveness of \name by answering the following research questions:
\begin{itemize}[leftmargin=0.6cm]
\item \textbf{RQ1} How accurately does \name recognize TTPs in decompiled functions?

\item \textbf{RQ2} How effective does \name recognize TTPs in malware binaries?
\item \textbf{RQ3} How does \name perform compared to baselines?
\item \textbf{RQ4} How do different components of \name contribute to overall performance? 
\end{itemize}

\subsection{Dataset construction}
\label{subsec:eval_dataset}
Table~\ref{tab:dataset} summarizes the malware datasets constructed for evaluation. Dataset~I provides function-level TTP labels, while Dataset~II provides binary-level TTP labels.

\noindent{\bf Dataset I. Function-Level Annotations} There is no publicly available dataset that labels individual binary functions with TTPs.
Existing malware reports typically document TTPs at the malware and rarely attribute them to specific functions.
Even when individual functions are discussed, reports often include only partial code snippets without calling context and do not provide the corresponding binaries, making precise function-level annotation infeasible.
To obtain reliable ground truth, we construct Dataset~I using malware source code, which exposes full semantic context and developer intent.
We collect 18 publicly leaked malware projects written in C/C++, compile each project into binaries, strip all symbols, and manually annotate source-level functions with their corresponding TTPs. Because compilation preserves function boundaries, each annotated source function can be mapped directly to its stripped binary counterpart. By comparing the number of source functions and binary functions (Columns 6 and 7), we observe that the function count in binaries increases by up to 40 times, primarily due to compiler-generated helper functions and linked library code. Some malware contains identical function implementations, and we remove such duplicated functions during dataset construction.
Dataset~I spans both Linux and Windows platforms, covers 12 malware families across 8 threat categories, and includes 42 distinct TTPs. In total, it includes 210 function–TTP pairs covering 122 functions.

\noindent{\bf Dataset II. Binary-level Annotations} 
To evaluate \name as an end-to-end system, we also include real-world malware binaries without available source code.  Dataset~II consists of eight binaries obtained from MalwareBazaar \cite{malwarebazaar} and Virusshare \cite{virusshare}. Among them, only two samples are accompanied by detailed human-written analysis reports that explicitly annotate TTPs. Such human-written reports are scarce in practice. For the remaining six binaries, although their malware families are documented in the MITRE ATT\&CK knowledge base, family-level TTP annotations are insufficient for labeling individual samples. Malware binaries within the same family may implement different subsets of TTPs, and without sample-specific analysis reports, MITRE family-level knowledge cannot be reliably used as ground-truth TTP labels. The available reports document TTPs at the malware level rather than at the function level, and some reported TTPs correspond to behaviors implemented in other execution stages or auxiliary binaries that are not included in the analyzed samples. We therefore filter out such TTPs and retain only those that are directly observable in the analyzed artifacts.

\begin{table*}[t]
    \setlength{\tabcolsep}{2pt}
    \caption{Dataset Statistics. The column 3,4,5 denote the category, family and project name for the malware. Column 6,7 reports the number of functions in the source code and in the binary file, separately. Column 8,9 report the number of unique TTPs and function-TTP pairs in the malware. ~\looseness=-1}
    \label{tab:dataset}
    \centering
    \scriptsize
\begin{adjustbox}{max width=\textwidth}
\begin{tabular}{cclllllll}
\toprule
 &
  Platform &
  Category &
  Family &
  Name &
  \# Src Func &
  \#Bin Func &
  \# TTPs &
  \# Pairs \\ 
\midrule
\multirow{17}{*}{Dataset I} &
  \multirow{8}{*}{Linux} &

Botnet & Gafgyt & Mirai.Linux.Lulz &
  138 & 310 & 9 & 23 \\

& & Botnet & Gafgyt & Renegade &
  81 & 244 & 1 & 1 \\

& & Botnet & Gafgyt & BallPit &
  53 & 234 & 4 & 8 \\

& & Botnet & Gafgyt & Botnet.Linux.LizardSquad &
  53 & 225 & 1 & 2 \\
& & Botnet & 
  Beastmode &
  Beastmode.d &
  135 &
  254 &
  11 &
  17 \\

& & Botnet & Mirai & Mirai.Linux.yakuza &
  50 & 241 & 1 & 1 \\

& & Backdoor & Lyceum & Linux.Lyceum.b &
  50 & 184 & 4 & 6 \\

& & Backdoor & BPFDoor & Linux.RedMenshenBPFDoor &
  25 & 199 & 7 & 14 \\
  
\cmidrule{2-9}

& \multirow{9}{*}{Windows} & Stealer & Predator the Thief & Win32.PredatorTheClipper &
  38 & 670 & 2 & 4 \\

& & Stealer & Predator the Thief & Win32.PredatorTheStealer.b &
  3737 & 7314 & 11 & 30 \\

& & Ransomware & Conti & Win32.Conti.c &
  505 & 5067 & 11 & 20 \\

& & banking-Trojan & Carberp & Anti\_rapport &
  27 & 279 & 5 & 12 \\

& & Tool & Generic & InjectDLL &
  9 & 218 & 2 & 3 \\

& & Tool & Generic & Portforw &
  33 & 159 & 5 & 7 \\

& & Tool & Generic & Schtasks &
  10 & 196 & 1 & 4 \\

&  & Point-of-sale (POS) & Alina & Alina &
  121 & 4474 & 16 & 28 \\

&  & Point-of-sale (POS) & Dexter & Dexter &
  61 & 247 & 11 & 17 \\

& & Spyware & Keylogger & Keylogger &
  25 & 1361 & 7 & 13 \\
  
\cmidrule{2-9}
& \multicolumn{4}{c}{Overall} &
  -- &  -- & 42 & 210 \\
\midrule
\multirow{8}{*}{Dataset II} &
 Linux & Trojan & Umbrellastand & Umbrellastand &
 -- &  763 & 3 & -- \\
\cmidrule{2-9}
& \multirow{7}{*}{Windows} & Trojan & Trojan:Win/Midie.Gen & DamascenedPeacock &
 -- &  797 & 4 & -- \\
      &  & Trojan & Zoxpng & Zoxpng &
 -- &  103 & -- & -- \\
 &  & Backdoor & AutoIt & AutoIt &
 -- &  469 & -- & -- \\
  &  & Backdoor & CozyDuke & Cozer &
 -- &  443 & -- & -- \\
   &  & Backdoor & Miniduke & Miniduke &
 -- &  452 & -- & -- \\
     &  & Backdoor & Zxshell & Zxshell &
 -- &  343 & -- & -- \\
    &  & Stealer & Strongpity & Strongpity &
 -- &  555 & -- & -- \\

\bottomrule
\end{tabular}
\end{adjustbox}

\end{table*}

\begin{table*}
\setlength{\tabcolsep}{2pt}
    \caption{Performance of \name on TTP recognition in functions compared with two basic prompts. Baseline~I and Baseline~II columns are discussed in RQ3. Column 1, 2, 3 denote the tactic name, technique id and technique name of the evaluated TTP. ~\looseness=-1}
    \centering

    \label{tab:eval_main}
\begin{adjustbox}{max width=\textwidth}
\begin{tabular}{lclccccccccccc}
\toprule
\multirow{2}{*}{\textbf{Tactic}} &
    \multicolumn{2}{c}{\textbf{Technique}} &
  \multicolumn{3}{c}{\textbf{Precision (\%)}} & 
  &
  \multicolumn{3}{c}{\textbf{Recall (\%)}} &
   &
  \multicolumn{3}{c}{\textbf{F1 (\%)}} \\
\cmidrule{2-3} \cmidrule{4-6} \cmidrule{8-10} \cmidrule{12-14}
 & ID & Name & Baseline I & Baseline II & \name & & Baseline I & Baseline II & \name & & Baseline I & Baseline II & \name \\
\midrule
Execution & T1059 & Command and Scripting Interpreter
& 100.00 & 100.00 & \textbf{100.00}
& & 100.00 & 100.00 & \textbf{100.00}
& & 100.00 & 100.00 & \textbf{100.00} \\
\midrule

\multirow{2}{*}{Persistence}
& T1547 & Boot or Logon Autostart Execution
& 42.86 & 50.00 & \textbf{62.50}
& & 60.00 & 100.00 & \textbf{100.00}
& & 50.00 & 66.67 & \textbf{76.92} \\
& T1112 & Modify Registry
& 100.00 & 100.00 & \textbf{100.00}
& & 70.00 & 90.00 & \textbf{100.00}
& & 82.35 & 94.74 & \textbf{100.00} \\
\midrule
\multirow{4}{*}{Defense Evasion}
& T1055 & Process Injection
& 85.71 & 100.00 & \textbf{100.00}
& & 100.00 & 100.00 & \textbf{100.00}
& & 92.31 & 100.00 & \textbf{100.00} \\
& T1036 & Masquerading
& 38.46 & 40.00 & \textbf{90.91}
& & 41.67 & 33.32 & \textbf{83.33}
& & 40.00 & 36.36 & \textbf{86.96} \\
& T1070 & Indicator Removal
& 46.15 & 57.14 & \textbf{100.00}
& & 60.00 & 80.00 & \textbf{70.00}
& & 52.17 & 66.67 & \textbf{82.35} \\
& T1562 & Impair Defenses
& 81.82 & 91.67 & \textbf{100.00}
& & 60.00 & 73.33 & \textbf{100.00}
& & 69.23 & 81.48 & \textbf{100.00} \\
\midrule
\multirow{3}{*}{Discovery}
& T1057 & Process Discovery
& 100.00 & 100.00 & \textbf{100.00}
& & 66.67 & 60.00 & \textbf{93.33}
& & 80.00 & 75.00 & \textbf{96.55} \\
& T1082 & System Information Discovery
& 100.00 & 90.91 & \textbf{93.33}
& & 64.29 & 71.43 & \textbf{100.00}
& & 78.26 & 80.00 & \textbf{96.55} \\
& T1046 & Network Service Discovery
& 50.00 & 66.67 & \textbf{75.00}
& & 83.33 & 66.67 & \textbf{100.00}
& & 62.50 & 66.67 & \textbf{85.71} \\
\midrule
Collection & T1074 & Data Staged
& 71.43 & 71.43 & \textbf{90.91}
& & 50.00 & 50.00 & \textbf{100.00}
& & 58.82 & 58.82 & \textbf{95.24} \\
\midrule
Command and Control & T1105 & Ingress Tool Transfer
& 85.71 & 100.00 & \textbf{100.00}
& & 85.71 & 100.00 & \textbf{100.00}
& & 85.71 & 100.00 & \textbf{100.00} \\
\midrule
Exfiltration & T1041 & Exfiltration Over C2 Channel
& 33.33 & 100.00 & \textbf{100.00}
& & 16.67 & 33.33 & \textbf{66.67}
& & 22.22 & 50.00 & \textbf{80.00} \\
\midrule
Impact & T1499 & Endpoint Denial of Service
& 76.47 & 92.31 & \textbf{92.86}
& & 100.00 & 92.31 & \textbf{100.00}
& & 86.67 & 92.31 & \textbf{96.30} \\
\midrule
\multicolumn{3}{c}{Average}
& 72.28 & 82.87 & \textbf{93.25}
& & 68.45 & 75.03 & \textbf{93.81}
& & 68.59 & 76.34 & \textbf{92.61} \\
\bottomrule
\end{tabular}
\end{adjustbox}
\end{table*}

\subsection{RQ1: Performance on Function-level TTP Recognition}

\noindent{\bf Setup and Metrics.}
To answer RQ1, we curate an evaluation dataset from malware in Dataset~I by selecting TTPs that appear at least 5 times, yielding 14 TTPs and 140 function–TTP pairs across 93 functions.
To balance the dataset, we sample 59 functions with no associated TTP labels and 29 functions labeled only with TTPs outside the selected TTP set from Dataset~I, resulting in a total of 181 functions. 
For each TTP, we evaluate \name on all 181 functions, resulting in 2{,}534 function-TTP binary classification instances. We report per-TTP precision, recall, and F1 score.

\noindent{\bf Results.} 
Columns labeled \textit{``\name\text{''}} in Table~\ref{tab:eval_main} report per-TTP performance on function-level TTP recognition.
\name achieves strong and consistent results across all evaluated TTPs, with an average precision of 93.25\%, recall of 93.81\%, and F1 score of 92.61\%. 
Especially, \name achieves 100\% precision and recall on five TTPs: T1059, T1112, T1055, T1562 and T1105. T1059 (Command and Scripting Interpreter), T1112 (Modify Registry) and T1105 (Ingress Tool Transfer) can be categorized as behavior-focused TTPs, which means detection of specific behaviors can infer the presence of TTP without much context and reasoning to understand the intent of the code. For example, some cases direct call the Windows Registry API functions RegCreateKeyEx,  RegSetValueEx, leading to the decision T1112. In contrast, T1562 and T1055 are harder to make decision, highlights the effectiveness of \name.
However, for T1547 (Boot or Logon Autostart Execution), the precision is a little bit lower. Some functions registers a scheduled task, which configures a new work item with logon trigger settings, with a logon trigger through Windows COM interfaces. \name missclassified it as T1547, but they better fall into the T1053 (Scheduled Task/Job). Though they serve the same purpose, but the implementation details are different. covers autostart mechanisms that hook into system initialization or user logon, like Registry Run Keys / Startup folders, Services created to auto-start and WMI Event subscriptions, etc.
These do achieve persistence at boot/logon, but the mechanism is distinct from scheduling functions. This subtle difference is not shown in the deffirentia criteria of the reasoning guideline.

\begin{figure}[t]
\centering
\includegraphics[width=0.75\columnwidth]{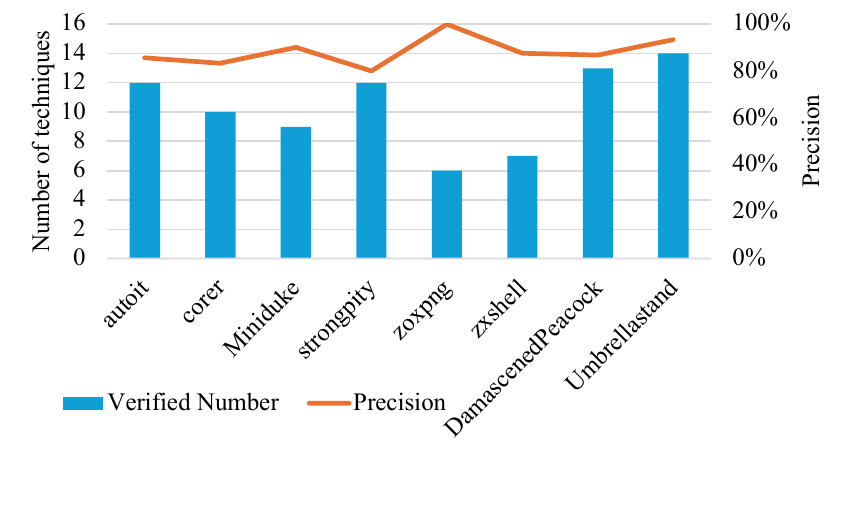}
\caption{Performance of TTP recognition in real-world malware binaries } 
\label{fig:binary_eval}
\vspace{-2mm}
\end{figure}

\begin{table*}[t]
\setlength{\tabcolsep}{2pt}
\caption{Binary-level malware TTP recognition results. Column~1 lists the malware name. Column~2 shows the number of ATT\&CK TTPs documented in expert reports, and Column~3 shows how many of these reported TTPs are predicted by \name. Column~4 is the report coverage. Column~5 reports the total number of TTPs predicted by \name, while Column~6 shows how many of these newly discovered TTPs are confirmed as true positives through human validation. Column~7 reports the  precision over all predicted TTPs. }

\label{tab:mal_acc}
\centering
\scriptsize
\begin{adjustbox}{max width=\linewidth}
\begin{tabular}{lcccccccc}
\toprule
Name &
\# Reported &
\# Covered &
Report Coverage (\%) &
\# Discovered &
\# True Positive &
Precision (\%) \\
\midrule

Umbrellastand &
3 &
3 &
100.0 &
15 &
14 &
93.3 \\

DamascenedPeacock &
4 &
3 &
75.0 &
15 &
13 &
86.7 & \\

\midrule
Overall &
7 &
6 &
85.7 &
24 &
21 &
90.0 \\
\bottomrule
\end{tabular}
\end{adjustbox}
\end{table*}

\begin{figure}[h]
\centering
\includegraphics[width=0.9\columnwidth]{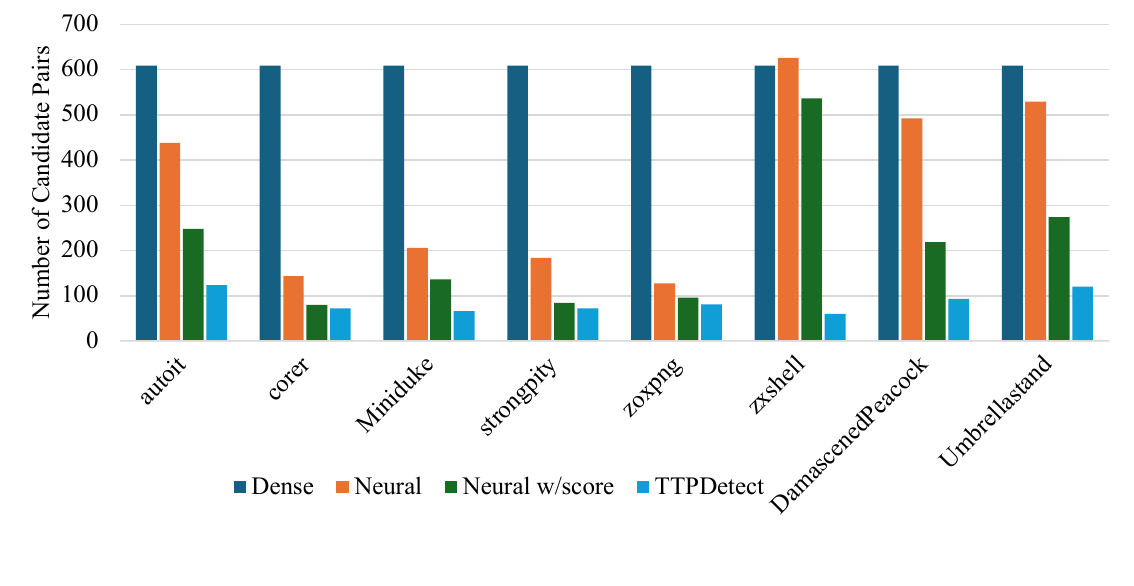}
\vspace{-2mm}
\caption{Number of function--TTP candidate pairs for analysis under different retrieval strategies.} 
\label{fig:eval_reduce}
\vspace{-3mm}
\end{figure}

\subsection{RQ2: Performance on Binary-level TTP Recognition}

\noindent{\bf Setup and Metrics.}
To answer RQ2, We evaluate \name for binary-level TTP recognition on Dataset~II, which contains all 203 TTPs.
For each malware binary, \name produces a set of predicted TTPs, which we evaluate using report coverage, precision, and number of function--TTP candidate pairs.
Report coverage measures the fraction of TTPs documented in expert reports that are recovered by \name.
Precision measures the fraction of TTPs predicted by \name that are verified as true positives. 
The number of function–TTP candidate pairs reflects the amount of downstream reasoning and analysis required.

\noindent{\bf Results.}
Fig.~\ref{fig:binary_eval} presents the binary-level TTP recognition results on Dataset~II. 
\name predicts 95 TTPs across all samples, of which 83 are confirmed as true positives, yielding an overall precision of 87.37\%. It doesn't report the report coverage because only two malware binaries have corresponding reports. Table~\ref{tab:mal_acc} reports the coverage. \name successfully recovers 6 of the 7 TTPs documented in expert reports, achieving a report coverage of 85.7\%. Beyond reported TTPs, \name identifies 24 previously unreported TTPs, 21 of which are validated as true positives through human analysis. These results demonstrate that \name is effective at discovering both known and previously undocumented TTPs at the binary level, providing additional insights that can support malware analysis. For example, in the case of damascened-peacock, \name identifies TTPs not noted in the report, such as T1497 (Virtualization/Sandbox Evasion), indicating that the malware may evade dynamic analysis and prompting analysts to adopt alternative analysis strategies.

Fig.~\ref{fig:eval_reduce} compares the number of function–TTP candidate pairs produced by different retrieval strategies. In addition to dense retrieval and the full \name pipeline, we consider two alternative retrieval variants. The first applies neural retrieval over all functions and is denoted as neural. The second further filters candidates based on the retrieval scores produced by neural retrieval and is denoted as neural w/ score.
Dense retrieval produces the largest candidate set. Neural retrieval reduces the number of candidates by 43.65\% compared to dense retrieval, but still introduces many function–TTP pairs. Adding score-based filtering further prunes the candidate space, achieving a 65.66\% reduction. In contrast, the full \name pipeline yields the smallest candidate set, reducing the number of candidates by 85.89\% relative to dense retrieval and substantially lowering the downstream reasoning effort.

\subsection{RQ3: Baseline Comparison}
\noindent{\bf Setup and Metrics.}
To answer RQ3, we compare \name against two basic prompting baselines. Baseline~I, derived from \cite{genttp, LADE}, applies a direct prompt to stripped decompiled functions without any customization for binary code. While \cite{cama} studies a different analysis task, it provides evidence that augmenting decompiled binaries with semantic information such as summaries is beneficial for LLM-based malware analysis. Accordingly, Baseline~II applies the same prompt as Baseline~I, but on stripped decompiled functions that are augmented with LLM generated function names and summaries.

\noindent{\bf Results.}
Table~\ref{tab:eval_main} (columns labeled \textit{``Baseline~I\text{''}} and \textit{``Baseline~II\text{''}}) reports function-level TTP recognition performance for the two baselines. 
Overall, \name substantially outperforms both baselines across all metrics, achieving higher average precision (93.25\% vs. 72.28\% for Baseline~I and 82.87\% for Baseline~II), recall (93.81\% vs. 68.45\% for Baseline~I and 75.03\% for Baseline~II), and F1 score (92.61\% vs. 68.59\% for Baseline~I and 76.34\% for Baseline~II).
Beyond these aggregate gains, \name consistently outperforms the baselines across nearly all individual ATT\&CK TTPs.
In particular, for T1036 (Masquerading), \name substantially outperforms both baselines, achieving an F1 score of 86.96\% versus 40.00\% for Baseline~I and 36.36\% for Baseline~II.

\subsection{RQ4: Ablation Study}

\begin{table*}
\setlength{\tabcolsep}{2pt}
\caption{Ablation study of different components. w/o Explorer columns reports the performance if we remove the explorer. w/o reason columns reports the performance if we remove the reasoning guideline. Best result in each group is highlighted in bold. ~\looseness=-1}
\centering

\label{tab:ablation}
\begin{adjustbox}{max width=\textwidth}
\begin{tabular}{clccccccccccc}
\toprule
\multicolumn{2}{c}{\textbf{Technique}} &
\multicolumn{3}{c}{\textbf{Precision (\%)}} & &
\multicolumn{3}{c}{\textbf{Recall (\%)}} & &
\multicolumn{3}{c}{\textbf{F1 (\%)}} \\
\cmidrule{1-2} \cmidrule{3-5} \cmidrule{7-9} \cmidrule{11-13}
ID & Name & 
w/o Explorer & w/o Reason & \name & &
w/o Explorer & w/o Reason & \name & &
w/o Explorer & w/o Reason & \name \\
\midrule
T1059 & Command and Scripting Interpreter &
100.00 & 100.00 & \textbf{100.00} & &
100.00 & 100.00 & \textbf{100.00} & &
100.00 & 100.00 & \textbf{100.00} \\
\midrule
T1547 & Boot or Logon Autostart Execution &
55.56 & 50.00 & \textbf{62.50} & &
100.00 & 80.00 & \textbf{100.00} & &
71.43 & 61.54 & \textbf{76.92} \\
T1112 & Modify Registry &
100.00 & 100.00 & \textbf{100.00} & &
90.00 & 90.00 & \textbf{100.00} & &
94.74 & 94.74 & \textbf{100.00} \\
\midrule
T1055 & Process Injection &
100.00 & 100.00 & \textbf{100.00} & &
100.00 & 100.00 & \textbf{100.00} & &
100.00 & 100.00 & \textbf{100.00} \\
T1036 & Masquerading &
62.50 & 46.15 & \textbf{90.91} & &
41.67 & \textbf{100.00} & 83.33 & &
50.00 & 63.16 & \textbf{86.96} \\
T1070 & Indicator Removal &
75.00 & 50.00 & \textbf{100.00} & &
60.00 & \textbf{90.00} & 70.00 & &
66.67 & 64.29 & \textbf{82.35} \\
T1562 & Impair Defenses &
91.67 & 93.33 & \textbf{100.00} & &
73.33 & 93.33 & \textbf{100.00} & &
81.48 & 93.33 & \textbf{100.00} \\
\midrule
T1057 & Process Discovery &
100.00 & 100.00 & \textbf{100.00} & &
80.00 & 86.67 & \textbf{93.33} & &
88.89 & 92.86 & \textbf{96.55} \\
T1082 & System Information Discovery &
100.00 & 80.00 & \textbf{93.33} & &
42.86 & 57.14 & \textbf{100.00} & &
60.00 & 66.67 & \textbf{96.55} \\
T1046 & Network Service Discovery &
66.67 & \textbf{83.33} & 75.00 & &
66.67 & 83.33 & \textbf{100.00} & &
66.67 & 83.33 & \textbf{85.71} \\
\midrule
T1074 & Data Staged &
85.71 & 69.23 & \textbf{90.91} & &
60.00 & 90.00 & \textbf{100.00} & &
70.59 & 78.26 & \textbf{95.24} \\
\midrule
T1105 & Ingress Tool Transfer &
100.00 & 87.50 & \textbf{100.00} & &
100.00 & 100.00 & \textbf{100.00} & &
100.00 & 93.33 & \textbf{100.00} \\
\midrule
T1041 & Exfiltration Over C2 Channel &
66.67 & 50.00 & \textbf{100.00} & &
33.33 & 50.00 & \textbf{66.67} & &
44.44 & 50.00 & \textbf{80.00} \\
\midrule
T1499 & Endpoint Denial of Service &
92.31 & 80.00 & \textbf{92.86} & &
92.31 & 92.31 & \textbf{100.00} & &
92.31 & 85.71 & \textbf{96.30} \\
\midrule
\multicolumn{2}{c}{Average} &
85.43 & 77.83 & \textbf{93.25} & &
74.30 & 86.63 & \textbf{93.81} & &
77.66 & 80.52 & \textbf{92.61} \\
\bottomrule
\end{tabular}
\end{adjustbox}
\end{table*}

\noindent{\bf Setup and Metrics.}
To assess the contribution of individual component, we conduct two ablation studies by removing either the Explorer or the Reasoning Guideline while keeping all other settings identical to those used in RQ1. We evaluate function-level ATT\&CK TTP recognition using the same per-TTP precision, recall, and F1 metrics.

\noindent{\bf Results.}
Table~\ref{tab:ablation} reports function-level TTP recognition performance under the ablated settings. Removing the Explorer affects recall more than precision. Average recall drops sharply from 93.81\% to 74.30\%, while precision decreases more modestly from 93.25\% to 85.43\%.
This indicates that, without the additional contextual evidence gathered by the Explorer, the model frequently fails to identify true TTP instances, resulting in many missed true positives.
In contrast, removing the Reasoning Guideline affects precision more than recall. Average precision decreases substantially from 93.25\% to 77.83\%, while recall drops more moderately from 93.81\% to 86.63\%. This reflects an increase in false positives due to the absence of structured reasoning constraints.
Overall, these results demonstrate that the Explorer and the Reasoning Guideline play complementary roles. The Explorer is essential for achieving high recall by expanding evidence coverage, while the Reasoning Guideline is critical for maintaining high precision by constraining the model’s inference process.

\section{Threats to Validity}

\textbf{Annotation Correctness.}
Our evaluation relies on manually annotated TTP labels derived from malware source code, which may be subject to human error or subjective interpretation. To mitigate this threat, two authors independently annotate the labels, provide explicit code-level evidence for assigning each TTP, and resolve disagreements through discussion.

\noindent\textbf{Generalization across LLMs.}
To assess whether \name generalizes across different LLMs, we conduct controlled experiments by replacing Claude~3.7~Sonnet in the analysis agent with \texttt{gpt-4\_1-2025-04-14} and \texttt{gpt-4o-2024-08-06}. \texttt{gpt-4o-2024-08-06} achieves 76.40\% precision and 74.5\% recall, while \texttt{gpt-4\_1-2025-04-14} achieves 89.50\% precision and 77.90\% recall. These results indicate that \name remains effective across different language models, although performance varies with the model’s reasoning capability. 

\noindent\textbf{Dataset Diversity \& Data Leakage.}
The main threat to external validity is the diversity of datasets used in our evaluation. 
The dataset we collected for evaluation may not fully represent the diversity of real-world malware.
To mitigate this threat, we cover different malware families on different platforms.
Another potential threat is data leakage during the pretraining of large language models. To mitigate this risk, we include Umbrellastand and DamascenedPeacock in our dataset, along with their corresponding reports. Both malware families first appeared after the training cutoff dates of the evaluated LLMs. The sample size is limited because such latest malware with publicly available reports is scarce.

\section{Related Work}
\subsection{Malware Analysis}
Traditional malware analysis relies on static analysis, dynamic execution, or handcrafted features to identify malicious behavior \cite{FlowDroid, Apposcopy, ASTROID, ndroid, betterpmp}. Learning-based approaches have also been applied to malware detection and family attribution \cite{Drebin, zhang_ccs14, TESSERACT, FeatureSmith}, but they primarily focus on coarse-grained identification or attribution and provide limited insight into the concrete behaviors implemented in code. More recently, large language models have been explored for malware analysis tasks such as malware detection and payload localization\cite{lamd, MalLoc, llm_malware_behavior}. 

\subsection{TTP Analysis}
Several studies aim to map malware behavior to the MITRE ATT\&CK framework using artifacts richer than raw binaries. One line of work relies on static, high-level artifacts such as cyber threat intelligence reports \cite{report_ttp1}, engineered static features \cite{droidttp}, executed code snippets \cite{LADE}, or interpreted malware packages with metadata \cite{genttp}. 
While effective in their respective settings, these approaches require information that is often unavailable in real-world analysis scenarios.
For example, GENTTP applies LLMs to generate TTPs from interpreted OSS malware by reasoning over package metadata and source code, which are typically not present in stripped binaries \cite{genttp}. 
Another line of work relies on dynamic information, including execution traces \cite{tracettp1} and system logs \cite{trec}, to infer ATT\&CK TTPs based on observed runtime behaviors. However, collecting such data requires heavyweight execution and monitoring, and is fragile in practice. Real-world malware frequently evades dynamic analysis through environment checks, delayed execution, or logic bombs, limiting the applicability of these methods\cite{forceexe}. 

As a result, existing methods are not directly applicable to stripped malware binaries that lack symbols, metadata, and execution traces. To the best of our knowledge, no prior work applies large language models to directly recognize MITRE ATT\&CK TTPs from stripped malware binaries.

\subsection{LLMs for Binary Analysis}

Recent work has explored the use of LLMs to recover semantic information from stripped binaries, primarily by generating function names or variable identifiers in an open-vocabulary manner \cite{symgen, gennm, ReSym}. 
While these approaches demonstrate that LLMs can reason effectively over decompiled code and infer meaningful program semantics, they are designed for program comprehension tasks and do not model malicious behavior.
In contrast, our work bridges stripped binary code with the MITRE ATT\&CK framework, enabling fine-grained, function-level TTP recognition directly and  behavior-centric analysis beyond semantic recovery.

\section{Conclusion}
We present \name, an LLM agent for function-level TTP recognition from stripped malware binaries. Unlike prior work that relies on high-level artifacts or execution traces, \name operates directly on binary code and focuses on static, behavior-centric analysis. To scale to large binaries, \name adopts a hybrid retrieval strategy that efficiently identifies analysis entry points, followed by demand-driven context retrieval and TTP-specific reasoning to support accurate function-level TTP identification.
Our evaluation shows that \name can accurately identify fine-grained ATT\&CK TTPs across diverse malware samples, demonstrating the feasibility of using LLMs for structured adversarial behavior modeling at the binary level.

\bibliographystyle{ACM-Reference-Format}
\bibliography{ref}

\appendix

\end{document}